\documentclass[preprint, showpacs, preprintnumbers, amsmath, amssymb]{revtex4}
\usepackage{graphicx}
\usepackage{dcolumn}
\usepackage{bm}

\begin{document}

\preprint{...}

\title{Topological aspect of black hole with Skyrme hair}
\author{Yi-Shi Duan}
\author{Xin-Hui Zhang}
\email{zhangxingh03@st.lzu.edu.cn}
\author{Li Zhao}
\affiliation{
Institute of Theoretical Physics, Lanzhou University \\
Lanzhou 730000, People's Republic of China}

%\date{\today}

\begin{abstract}

Based on the $\phi$-mapping topological current theory, we show that the
presence of the black hole leaves fractional baryon charge outside the
horizon in the Einstein-Skyrme theory. A topological current is derived
from the Einstein-Skyrme system, which corresponds to the monopoles around
the black hole. The branch process (splitting, merging and intersection) is
simply discussed during the evolution of the monopoles.

\end{abstract}
\maketitle

\section{introduction}

The famous conjecture ``Black holes have no hair" is that black holes are
uniquely specified by the conserved quantities: mass, angular momentum, and
charge \cite{1}. Investigating these ``no-hair" theorems, however, shows
that while powerful, they are not omnipotent. It was pointed out by Bowick
\emph{et. al}. \cite{2} that there exists a family of Schwarzschild
black-hole solutions to the Einstein-axion equations labelled by a
conserved topological charge. Such black holes could be said to be carrying
axion hair. It was then rapidly realized that the same fractional charge
that could give rise to enhancement of proton decay catalysis
\cite{Gerbert1989,Alford1989,KraussPRL1989}. The full ramifications of this
type of quantum hair have been most eloquently argued by Coleman \emph{et.
al}. \cite{11,12}.

The Skyrme model is an effective meson theory \cite{skyrme1961,skyrme1962}
where the baryons are identified with topological solitons, so-called
skyrmions. The baryon number $B$ corresponds to the topological charge
\cite{witten}. Thus the model gives a unified description of hadron physics
in terms of meson fields \cite{norikoprd}. The advantage of the Skyrme
model to more realistic nucleon models is that it is straightforward to
couple to gravity, that is, the Einstein-Skyrme model \cite{norikoarxiv}.
The Einstein-Skyrme model was firstly studied by Luckock and Moss
\cite{n14} where the Schwarzschile black hole with Skyrme hair was
obtained. This is another counter example of the no-hair conjecture for
black holes \cite{1}. The full Einstein-Skyrme system was solved later to
obtain spherically symmetric black holes with Skyrme hair \cite{n16} and
regular gravitating skyrmions \cite{n17}.

Here we shall be particular interested in the Einstein-Skyrme model and the
hair of the black hole. We reexamine some work in detail using the
$\phi$-mapping topological current theory and present there leaves
fractional baryon charges outside the horizon of black hole in the
Einstein-Skyrme model. Meanwhile, in this system there also exist
monopoles, whose quantum number is determined by Hopf indices and Brouwer
degrees. Thus, in some sense, such black holes could be said to be carrying
Skyrme hair and monopole hair at the same time.

This paper is arranged as follows. In Sec. \ref{sec2}, we research the
topological current in the Einstein-Skyrme model and reveal that it can be
decomposed into two parts. One is the usual baryon number current, i.e.,
the skyrmion current. The other is the monopole current. In Sec.
\ref{sec3}, We study the baryon number current and point out that baryon
number may be fractional outside the horizon. In Sec. \ref{sec4}, using the
$\phi$-mapping topological current theory, we investigate the inner
topological structure of the monopole current and simply discuss the branch
process during the evolution of the monopoles. The conclusion of this paper
is given in Sec. \ref{sec5}.

\section{Topological current in Einstein-Skyrme model}\label{sec2}

%\subsection{Baryon number current and monopole current}

We write the field $U$ in SU(2) space in a general form
\begin{eqnarray}
U&=&e^{inF(\varphi)}=\cos{F}+in\sin{F}, \label{uu}\\
n&=&n^a\sigma_a,\;\;\;\;\;n^a=\frac{\phi^a}{\|\phi\|},\label{n}
\end{eqnarray}
where $\phi^a\;(a=1,2,3)$ are three fundamental functions with space-time
coordinates $x^\mu$ and $\sigma^a$ are pauli matrices. Then the
Einstein-Skyrme model is denoted by the Lagrangian
\begin{eqnarray}
L&=&L_G+L_S\\\nonumber
&=&\frac{R}{16\pi{G}}+\frac{F_{\pi}}{16}g^{\mu\nu}\textrm{tr}(L_\mu{L}_\nu)+
\frac{1}{32a^2}g^{\mu\nu}g^{\rho\sigma}\textrm{tr}([L_\mu,L_\rho][L_\nu,L_\sigma]),
\end{eqnarray}
in which $R$ is the scalar curvature, $G$ is the Newton's constant and
$L_{\mu}=U^+\partial_{\mu}U$. It is well known that the Einstein-Skyrme
model has black hole solutions \cite{n14,n16,n17}. Here we leave out the
concrete process of solving the Einstein equations and directly explore the
topological properties of black hole solutions in the Einstein-Skyrme
system. Correspondingly, we shall impose the spherically symmetric ansatz
on the metric
\begin{equation}
ds^2=-N^2(r)C(r)dt^2+\frac{1}{C(r)}dr^2+r^2d\Omega^2,
\end{equation}
where $C(r)$ is defined as follows:
\begin{equation}
C(r)=1-\frac{2Gm(r)}{r}.
\end{equation}
At the horizon $r=r_h$, we have $C(r_h)=0$, that is, $m(r_h)=r_h/2G$. It
should be noted that in the presence of  black hole, the matter field is
defined only outside the horizon and therefore the integral over the space
is performed from the horizon to infinity. The covariant topological
current in this model is defined by \cite{skyrme1961,skyrme1962}
\begin{equation}\label{bj}
B^\mu=\frac{\epsilon^{\mu\nu\lambda\rho}}{24\pi^2}\frac{1}{\sqrt{-g}}
\textrm{tr}[U^{-1}(\partial_{\nu}U)U^{-1}(\partial_{\lambda}U)U^{-1}\partial_{\rho}U],
\end{equation}
whose $0$-th component can be expressed as
\begin{equation}\label{bu}
B(U)=\frac{\epsilon^{0ijk}}{24\pi^{2}}\frac{1}{\sqrt{-g}}\textrm{tr}[U^{-1}
(\partial_{i}U)U^{-1}(\partial_{j}U)U^{-1}\partial_kU],\;\;i,j,k=1,2,3,
\end{equation}
which is just the topological current density. $B(U)$ defined by Eq.
(\ref{bu}) is additive
\begin{equation}
  B(U_1U_2)=B(U_1)+B(U_2).
\end{equation}
Substituting Eq. (\ref{n}) into Eq. (\ref{bj}) and taking notice of the
relations: $n^2=1$ and $n\partial_\mu{n}+\partial_{\mu}nn=0$, we find
$B^{\mu}$ is composed of two parts
\begin{equation}
B^\mu=B^\mu_1+B^\mu_2,
\end{equation}
where
\begin{equation}\label{b1}
B_1^\mu=\frac{1}{4\pi^2}\frac{1}{\sqrt{-g}}\epsilon^{\mu\nu\lambda\rho}
\epsilon_{abc}\sin^2F\partial_\nu{F}n^a\partial_\lambda{n}^b\partial_\rho{n}^c,
\end{equation}
and
\begin{equation}\label{b2}
B^\mu_2=\frac{1}{12\pi^2}\frac{1}{\sqrt{-g}}\epsilon^{\mu\nu\lambda\rho}\epsilon_{abc}
\sin^3F\cos{F}\partial_{\nu}n^a\partial_\lambda{n}^b\partial_\rho{n}^c.
\end{equation}
We shall see later that the current $B^\mu_1$ denoted by Eq. (\ref{b1}) is
identical with the baryon number current, i.e. the skyrmion current
\cite{baryoncurrent}, while the current $B^\mu_2$ denoted by Eq. (\ref{b2})
corresponds to the monopoles with charge $ g=({2}/{3\pi})\sin^3F\cos{F}.$

\section{Fractional baryon charges}\label{sec3}

 In this section, we discuss the baryon number current $B^\mu_1$.
From Eq. (\ref{b1}) we find the baryon number density
\begin{equation}\label{rho}
\rho(x)=\frac{1}{\sqrt{-g}}\rho(\phi)D(\frac{\phi}{x}),
\end{equation}
where
\begin{eqnarray}
\rho(\phi)=\frac{1}{2\pi^2}\frac{\sin^2{F}}{\phi^2}\frac{dF}{d\phi},\;\;\;
D(\frac{\phi}{x})=\frac{1}{2}\epsilon^{ijk}\epsilon_{abc}
\partial_{i}n^a\partial_j{n}^b\partial_k{n}^c.
\end{eqnarray}
Then the baryon number can be expressed as
\begin{equation}\label{brho}
  B_1=\int_V\rho(x)\sqrt{-g}d^3x=\int_T\rho(\phi)d^3\phi.
\end{equation}
The above integral shows that the fundamental field $\phi$ maps $V$ into
$T$ in iso-space. We consider the case that while the vector $\vec{x}$
covers the region $V$ once, function $\phi$ covers the corresponding
elementary region $T_0$ $N$ times (i.e. $T=NT_0$) and impose the boundary
conditions on the profile function
\begin{equation}
F(\infty)=0,\;\;F(r_h)=F_h,
\end{equation}
which determines the value at the horizon $r_h$. Then from Eq. (\ref{rho})
and Eq. (\ref{brho}), we obtain the baryon number
\begin{equation}
B_1=\frac{2N}{\pi}\int^{F_{h}}_0\sin^2{F}dF=\frac{N}{2\pi}(2F_{h}-\sin{2F_{h}}).
\end{equation}
If $F_h=\pi$, the baryon number is still an integer. This configuration
represents a proton tightly bound to the black hole. On the other hand, if
$F_{h}<\pi$, the baryon number may not be an integer and the skyrmion can
carry fractional baryon number. This configuration will be interpreted as a
proton partially swallowed by the black hole.

\section{Monopoles outside the horizon}\label{sec4}

\subsection{The generation of the monopole}\label{sec4.1}
Generally speaking, one explores the topological feature of the
Einstein-Skyrme model only by the baryon current $B_1^\mu$. In Sec.
\ref{sec2}, we obtain another current $B_2^\mu$, i.e. monopole topological
current, which contributes rich properties to the black hole with skyrmion
hair. In this subsection, using the $\phi$-mapping topological current
theory, we investigate the inner topological structure of $B^\mu_2$. Eq.
(\ref{b2}) can be rewritten as
\begin{equation}\label{b22}
B^\mu_2=\frac{g_0}{8\pi}\frac{1}{\sqrt{-g}}\epsilon^{\mu\nu\lambda\rho}\epsilon_{abc}
\partial_{\nu}n^a\partial_\lambda{n}^b\partial_\rho{n}^c,
\end{equation}
where $g_0=({2}/{3\pi})\sin^3F\cos{F}$. Taking account of Eq. (\ref{n}) and
using
$\partial_\mu{n}^a=\phi^a\partial_\mu({1/\|\phi\|})+(1/\|\phi\|)\partial_\mu\phi^a$,
Eq. (\ref{b22}) can be expressed as
\begin{equation}
B^\mu_2=\frac{g_0}{8\pi}\frac{1}{\sqrt{-g}}\epsilon^{\mu\nu\lambda\rho}\epsilon_{abc}
\frac{\partial}{\partial\phi^l}\frac{\partial}{\partial\phi^a}(\frac{1}{\|\phi\|})
\partial_{\nu}\phi^l\partial_{\lambda}\phi^b\partial_{\rho}\phi^c.
\end{equation}
If we define Jacobian
$\epsilon^{lbc}D^\mu(\frac{\phi}{x})=\epsilon^{\mu\nu\lambda\rho}\partial_{\nu}\phi^l
\partial_{\lambda}\phi^b\partial_{\rho}\phi^c,$
and make use of the Green function relation in ${\phi}$ space:
$\frac{\partial}{\partial\phi^a}\frac{\partial}{\partial\phi^a}(\frac{1}{\|\phi\|})=
-4\pi\delta(\vec{\phi})$, we obtain the $\delta$-function like $B^\mu_2$
\cite{phi1998,phi1999,phi2003}
\begin{equation}\label{bu2}
B^\mu_2=\frac{1}{\sqrt{-g}}g_0\delta(\vec{\phi})D^\mu(\frac{\phi}{x}).
\end{equation}
The expression of Eq. (\ref{bu2}) provides an important conclusion:
\begin{equation}
B^\mu_2\left\{
\begin{array}{l}
=0,\;\textrm{if and only if}\;{\vec{\phi}}\neq 0. \\
\neq 0,\;\textrm{if and only if}\;{\vec{\phi}}=0.
\end{array}
\right.
\end{equation}
So it is necessary to study the zero points of ${\vec{\phi}}$ to determine
the nonzero solutions of $B^\mu_2$. The implicit function theory
\cite{implicite} shows that under the regular condition
\begin{equation}
D^\mu({\phi} /x)\neq 0,
\end{equation}
the general solutions of
\begin{equation}\label{phi}
{\phi}^1(t,x^1,x^2,x^3)=0,\;\;{\phi}^2(t,x^1,x^2,x^3)=0,\;\;\phi^3(t,x^1,x^2,x^3)=0
\end{equation}
can be expressed as
\begin{equation}\label{xxx}
x^{1}=x^{1}_{k}(t),\;\;x^{2}=x^{2}_{k}(t),\;\;x^3=x^3_k(t),
\end{equation}
which represent the world lines of $N$ moving isolated singular points
$\vec{x}_{k}$ $(k=1,2,3\cdots{N})$. These singular solutions are just the
monopoles located at the zero point of field $\vec\phi$ outside the horizon
of the black holes. Then questions are raised naturally: what are the
topological charges of the monopoles and how to obtain the inner structure
of $B^\mu_2$. Now, we investigate the topological charges of the monopoles
and their quantization. Let $M_k$ be a neighborhood of $\vec{x}_k$ with
boundary $\partial{M}_k$ satisfying $\vec{x}_k\nsubseteqq{\partial{M}_k}$,
$M_i\bigcap{M}_k=\varnothing$. Then the generalized winding number $W_k$ of
$n^a{(\vec{x})}$ at $\vec{x}_k$ can be defined by the Gauss map
\cite{ying24} $n: \partial{M}_k\rightarrow{S^2}$,
\begin{equation}
W_k=\frac{1}{8\pi}\int_{\partial{M_k}}n^*(\epsilon_{abc}n^adn^b\wedge{dn^c}),
\end{equation}
where $n^*$ is the pull back of map $n$. The generalized winding number is
a topological invariant and is also called the degree of Gauss map. It is
well known that $W_k$ are corresponding to the second homotopy group
$\pi_2(S^2)=Z$ ( the set of integers). Using the Stokes' theorem in
exterior differential form and the result in Eq. (\ref{bu2}), we get
\begin{equation}\label{wk}
W_k=\int_{M_k}\delta(\vec\phi)D(\frac{\phi}{x})d^3x.
\end{equation}
In order to explore the inner topological structure of $B^\mu_2$, one can
expand the $\delta(\vec\phi)$ as:
\begin{equation}\label{ck}
\delta(\vec\phi)=\sum^N_{k=1}c_k\delta(\vec{x}-\vec{x}_k),
\end{equation}
where the coefficients $c_k$ must be positive, i.e., $c_k=|c_k|$.
Substituting Eq. (\ref{ck}) into Eq. (\ref{wk}) and calculating the
integral, we obtain an expression for $c_k$,
\begin{equation}
c_k=\frac{|W_k|}{|D(\phi/x)|_{\vec{x}=\vec{x}_k}}.
\end{equation}
Then $\delta(\vec{\phi})$ can be reexpressed as
\begin{equation}\label{delta}
\delta(\vec\phi)=\sum^N_{k=1}\frac{\beta_k}{|D(\frac{\phi}{x})|_{\vec{x}=\vec{x}_k}}
\delta(\vec{x}-\vec{x}_k),
\end{equation}
in which the positive integer $\beta_{k}=|W_k|$ is the Hopf index of
$\phi$-mapping, which means that when $\vec{x}$ covers the neighborhood of
the zero point $\vec{x}_{k}$ once, the vector field $\vec\phi$ covers the
corresponding region in $\phi$ space $\beta_{k}$ times. It can be proved
from Eq. (\ref{xxx}) that the velocity of the $k$th zero is determined by
\cite{liuxin12}
\begin{equation}\label{velocity}
V^\mu=\frac{dx^\mu_k}{dt}=\left.\frac{D^\mu(\frac{\phi}{x})}{D^0(\frac{\phi}{x})}\right|_{\vec{x}=\vec{x}_k},
\end{equation}
where $D^0(\frac{\phi}{x})$ is the usual Jacobian $D(\frac{\phi}{x})$. Then
substituting Eqs. (\ref{velocity}) and (\ref{delta}) into Eq. (\ref{bu2}),
we get the dynamic form of the current $B^\mu_2$
\begin{equation}
B^\mu_2=\frac{1}{\sqrt{-g}}g_0\sum^N_{k=1}\beta_k\eta_k
\delta(\vec{x}-\vec{x}_k(t))\frac{dx^\mu_k}{dt}=\rho{V}^\mu
\end{equation}
and the topological charge
\begin{equation}
B_2=\int_V\rho\sqrt{-g}{d^3}x=\int_Vg_0\sum^N_{k=1}\beta_k
\eta_k\delta(\vec{x}-\vec{x}_k(t))d^3x =g_0\sum^N_{k=1}W_k,
\end{equation}
where the integral over the space is  from the horizon to infinity,
$W_k=\beta_k \eta_k$ is the winding number and
$\eta_k=\textrm{sgn}D(\phi/x)|_{\vec{x}_k}=\pm1$ is called the Brouwer
degree of the $\phi$-mapping at $\vec{x}_k$. From the above discussions, we
see that:

(i) The monopoles are generated from the zeros of the field $\vec\phi$
outside the horizon of the black holes.

(ii) The monopoles are topologically quantized in the unit of the basic
charge $g_0$ and the topological quantum numbers are determined by Hopf
indices $\beta_k$ and Brouwer degrees $\eta_k$ of the $\phi$-mapping.

(iii) The Brouwer degree $\eta_k=+1$ corresponds to the monopoles, while
$\eta_k=-1$ corresponds to the anti-monopoles.

\subsection{The evolution of the monopoles}\label{sec4.2}

In Sec. \ref{sec4.1}, we have used the regular condition
$D^\mu(\phi/x)\neq0$. When this condition fails, branch process will occur.
Here we discuss branch process during the evolution of monopoles. It is
known that the velocity of the monopoles is given by
\begin{equation}\label{volicity2}
V^\mu=\frac{dx^\mu}{dt}=\frac{D^\mu(\phi/x)}{D^0(\phi/x)},
\end{equation}
where
\begin{eqnarray}
D^0(\phi/x)=\epsilon_{abc}\partial_{1}\phi^a\partial_{2}\phi^b\partial_{3}\phi^c,\;\;\;
D^1(\phi/x)=\epsilon_{abc}\partial_{2}\phi^a\partial_{3}\phi^b\partial_{0}\phi^c,\nonumber \\
D^2(\phi/x)=\epsilon_{abc}\partial_{3}\phi^a\partial_{0}\phi^b\partial_{1}\phi^c,\;\;\;
D^3(\phi/x)=\epsilon_{abc}\partial_{0}\phi^a\partial_{1}\phi^b\partial_{2}\phi^c.\nonumber
\end{eqnarray}
From Eq. (\ref{volicity2}), it is obvious that when
\begin{equation}
D^0(\phi/x)=0,
\end{equation}
at the very point $(t^*,\vec{x}^*)$, the velocity
\begin{eqnarray}
\frac{dx^1}{dt}=\left.\frac{D^1(\phi/x)}{D^0(\phi/x)}\right|_{(t^*,\vec{x}^*)},\;\;\;
\frac{dx^2}{dt}=\left.\frac{D^2(\phi/x)}{D^0(\phi/x)}\right|_{(t^*,\vec{x}^*)},
\frac{dx^3}{dt}=\left.\frac{D^3(\phi/x)}{D^0(\phi/x)}\right|_{(t^*,\vec{x}^*)}
\end{eqnarray}
are not unique in the neighborhood of $(t^*,\vec{x}^*)$. This very point
$(t^*,\vec{x}^*)$ is called the bifurcation point. Without loss of
generality, we discuss only the branch of the velocity component
$(dx^1/dt)$ at $(t^*,\vec{x}^*)$. It is known that the Taylor expansion of
the solution of Eq. (\ref{xxx}) in the neighborhood of $(t^*,\vec{x}^*)$
can generally be expressed as
\begin{equation}
A(x^1-x^{1*})^2+2B(x^1-x^{1*})(t-t^*)+C(t-t^*)^2+\cdot\cdot\cdot=0,
\end{equation}
where $A$, $B$ and $C$ are three constants. Then the above Taylor expansion
leads to
\begin{equation}\label{abc}
A\left(\frac{dx^1}{dt}\right)^2+2B\frac{dx^1}{dt}+C=0\;\;\;(A\neq0).
\end{equation}
The solutions of Eq. (\ref{abc}) give different motion directions of the
zero point at the bifurcation point. There are two possible cases:

(i) For $\Delta=4(B^2-AC)=0$, from Eq. (\ref{abc}) we get only one motion
direction of the zero point at the bifurcation point:
$(dx^1/dt)|_{1,2}=-B/A$, which includes three sub-cases:

(a) One point defect splits into two points;

(b) Two points merge into one;

(c) Two point defects tangentially intersect at the bifurcation point.

(ii) For $\Delta=4(B^2-AC)>0$, from Eq. (\ref{abc}) we get two different
motion directions of the zero point:
$(dx^1/dt)|_{1,2}=(-B\pm\sqrt{B^2-AC})/A$. This is the intersection of two
point defects, which means that the two defects meet
and then depart at the bifurcation point.\\
\indent In both cases (i) and (ii),  we know that the sum of the
topological charges of final defects must be equal to that of the initial
defects at the bifurcation point.

\section{Conclusion}\label{sec5}

In this paper, we investigate the topological aspect of black hole with
Skyrme hair in Einstein-Skyrme model. It is revealed that the baryon number
may be fractional outside the horizon and there exist monopoles, whose
topological charges are topologically quantized by Hopf indices and Brower
degrees. Starting with the Einstein-Skyrme model, which supports black hole
solutions, we discuss the baryon number current $B_1^\mu$ outside the
horizon of the black hole and come to the conclusion: if the baryon number
is still an integer, this configuration represents a proton tightly bound
to the black hole; On the other hand, if the baryon number is not an
integer, i.e. the skyrmion carries fractional baryon number, this
configuration will be interpreted as a proton partially swallowed by the
black hole. Meanwhile we find there is a topological current $B_2^\mu$
accompanied by the baryon number current $B_1^\mu$. We research the current
$B_2^\mu$ in detail and point out that it corresponds to monopoles outside
the horizon. By introducing the $\phi$-mapping topological current theory,
we see that the zeros of the field $\vec\phi$ are just the sources of the
monopoles, and the monopoles are topologically quantized in the unit of the
basic charge $g_0$ and the topological quantum numbers are determined by
the Hopf indices $\beta_k$ and Brower degrees $\eta_k$. Finally, we simply
discuss the branch process (splitting, merging and intersection) when the
condition $D^\mu(\phi/x)\neq0$ fails. The topological charges are preserved
in the branch process during the evolution of the monopoles.

\section{acknowledgment}

It is a great pleasure to thank Doctor T. Y. Si  and Doctor Y. X. Liu for
numerous fruitful discussions.  This work was supported by the National
Natural Science Foundation of China and the Doctoral Foundation of the
People's Republic of China.

\end{document}